\documentclass[iop,apjl,12pt]{emulateapj}
\usepackage{times}
\usepackage{apjfonts}
\usepackage{epsfig}
\usepackage{natbib}
\usepackage{amsfonts}
\usepackage{amsmath}
\usepackage{multirow}
\usepackage{enumerate}
\usepackage[usenames]{color}
\usepackage[plainpages=false, colorlinks=true, anchorcolor=lblui, linkcolor=blue, citecolor=black, urlcolor=blue, bookmarks=false]{hyperref}
\newcommand{\newacronym}[3]{%
 \newcommand{#1}[1]{#3##1 (#2##1)%
 \renewcommand{#1}[1]{#2####1}}%
}

\newcommand{\fullhi}{s27FH}
\newcommand{\fulllo}{s27FL}
\newcommand{\octhi}{s27OH}  
\newcommand{\octlo}{s27OL} 
\newacronym{\SN}{SN}{supernova}

\newacronym{\NSNS}{NSNS}{neutron star--neutron star}
\newacronym{\SPH}{SPH}{smoothed particle hydrodynamics}
\newacronym{\BH}{BH}{black hole}
\newacronym{\NS}{NS}{neutron star}
\newacronym{\sGRB}{sGRB}{short gamma ray burst}
\newacronym{\EOS}{EOS}{equation of state}
\newacronym{\NSE}{NSE}{nuclear statistical equilibrium}

\newcommand\cqg{{Class. Quantum Grav.}}

\def\Dwa{$\,$\uppercase\expandafter{\romannumeral5}$\,$}

\def\sless{\lower2pt\hbox{$\buildrel {\scriptstyle <}
   \over {\scriptstyle\sim}$}}

\def\sgreat{\lower2pt\hbox{$\buildrel {\scriptstyle >}
   \over {\scriptstyle\sim}$}}
\def\sharpnull#1{}

\newcommand{\newtext}[1]{#1}

\newcommand{\shortauth}{Roberts \emph{et al.}}
\newcommand{\slugcom}{Draft version - \today}
\slugcomment{\slugcom}

\lefthead{\sc \footnotesize \slugcom \hfill \shortauth}
\righthead{\sc \footnotesize \slugcom \hfill \shortauth}

\begin{document}
\slugcomment{Draft version \today.}

\title{General Relativistic Three-Dimensional Multi-Group Neutrino Radiation-Hydrodynamics Simulations of Core-Collapse Supernovae}

\author{Luke F. Roberts\altaffilmark{1,+}}
\author{Christian D. Ott\altaffilmark{2,1}}
\author{Roland Haas\altaffilmark{3,1}}
\author{Evan P. O'Connor\altaffilmark{4,++}}
\author{Peter Diener\altaffilmark{5,6}}
\author{Erik Schnetter\altaffilmark{7,8,5}}
\altaffiltext{1}{TAPIR, Walter Burke Institute for Theoretical Physics,
  Mailcode 350-17,
  California Institute of Technology, Pasadena, CA 91125, USA, 
  lroberts@tapir.caltech.edu}
\altaffiltext{2}{Yukawa Institute for Theoretical Physics, Kyoto University, Kyoto, Japan}
\altaffiltext{3}{Max Planck Institut f\"ur Gravitationsphysik, Potsdam, Germany.}
\altaffiltext{4}{Department of Physics and Astronomy, North Carolina State University, NC, USA.}
\altaffiltext{5}{Center for Computation \& Technology, Louisiana State University, Baton Rouge, USA.}
\altaffiltext{6}{Department of Physics \& Astronomy, Louisiana State University, Baton Rouge, USA.}
\altaffiltext{7}{Perimeter Institute for Theoretical Physics, Waterloo, ON, Canada.}
\altaffiltext{8}{Department of Physics, University of Guelph, Guelph, ON, Canada.}
\altaffiltext{+}{NASA Einstein Fellow}
\altaffiltext{++}{NASA Hubble Fellow}

\begin{abstract}
We report on a set of long-term general-relativistic three-dimensional
(3D) multi-group (energy-dependent) neutrino-radiation hydrodynamics simulations
of core-collapse supernovae. We employ a full 3D two-moment scheme with the
local M1 closure, three neutrino species, and 12 energy groups per species. With
this, we follow the post-core-bounce evolution of the core of a nonrotating
$27$-$M_\odot$ progenitor in full unconstrained 3D and in octant symmetry for
$\gtrsim$$ 380\,\mathrm{ms}$. We find the development of an
asymmetric runaway explosion in our unconstrained simulation. We test the
resolution dependence of our results and, in agreement with previous work, find
that low resolution artificially aids explosion and leads to an earlier runaway
expansion of the shock. At low resolution, the octant and full 3D dynamics are
qualitatively very similar, but at high resolution, only the full 3D simulation
exhibits the onset of explosion.  
\end{abstract}


\section{Introduction}
Although it has been studied for many decades, the mechanism driving
core-collapse supernova explosions (CCSNe) is still uncertain and an
area of active research \citep[e.g.][]{janka:12a,burrows:13a}.  The
delayed neutrino mechanism \citep{bethewilson:85}, in combination with
multi-dimensional fluid instabilities, seems to be the most promising
mechanism driving garden-variety CCSNe. However, it cannot deliver the
explosion energies seen in some extreme CCSNe (hypernovae). Another
mechanism must most likely be at work in these events, possibly
relying on rotation and magnetic fields (e.g.,
\citealt{burrows:07b,moesta:14b,moesta:15}).

The hydrodynamic shock formed at core bounce stalls due to energy loss
to dissociation of heavy nuclei and to neutrinos.  The delayed
neutrino-heating model for CCSNe posits that a small fraction of the
neutrinos emitted from near the protoneutron star are absorbed near
the stalled shock, thereby depositing enough energy to reinvigorate
the shock's outward progress. This shock revival must occur within
few~$100\,\mathrm{ms}$ to $\sim$$1-2\,\mathrm{s}$ of core bounce to
avoid black hole formation or a top-heavy neutron star mass
distribution \citep{oconnor:11}. Since the neutrino mechanism
strongly depends on how efficiently energy is transported by neutrinos
from near the protoneutron star to the region just behind the shock
and on how this energy deposition effects the hydrodynamic evolution
near the shock, an accurate treatment of hydrodynamics and
non-equilibrium neutrino transport are key requirements for simulating
CCSNe.

Imposing symmetries on simulations of CCSNe can have significant
consequences for their ev\cite[e.g.][]{hanke:12,murphy:13,couch:13b}.  Detailed spherically
symmetric (1D) simulations do not explode \citep{liebendoerfer:01b},
except when very particular low mass progenitor models are used
\citep[e.g.][]{fischer:10, huedepohl:10}.  Multiple simulations
including energy-dependent (multi-group) neutrino transport and
imposing axial symmetry (2D) do exhibit explosions
\citep{mueller:12a,mueller:12b,bruenn:13,bruenn:16}, some do not
\citep{dolence:15}.  Interestingly, the first
simulations including ray-by-ray neutrino
transport\footnote{Ray-by-ray solves individual 1D transport problems
  along radial rays that are coupled via lateral advection terms.}
without symmetries imposed (3D) on the hydrodynamics did not find
explosions in models that exploded when axisymmetry was assumed
\citep{hanke:13}. \cite{melson:15b} showed that in 3D simulations that
are close to the threshold of explosion, modified neutrino interaction
physics can facilitate explosion. \cite{lentz:15} carried out 1D, 2D,
and 3D simulations, using a ray-by-ray multi-group flux-limited
diffusion approximation to neutrino transport. They found explosions
in 2D and 3D, with an earlier onset of explosion in 2D.

The differences between 2D and 3D are likely due to the evolution of
postshock hydrodynamic instabilities, namely the standing accretion
shock instability (SASI) and turbulent convection, when different
symmetries are enforced \citep{couch:13b,couch:15a}.  Clearly, these
non-radial instabilities are completely suppressed in spherical
symmetry.  There are also significant differences between 2D and full
3D for both of these instabilities.  Azimuthal modes are suppressed in
axisymmetry which has consequences for the evolution of the SASI
\citep[e.g.,][]{iwakami:08}.  Additionally, it is well known that
the properties of two-dimensional turbulence differ significantly from
those of three-dimensional turbulence \citep{kraichnan:67, hanke:12}. In
particular, 2D turbulence, because of the conservation of enstrophy in
2D, exhibits an inverse cascade. This inverse cascade transfers
kinetic energy to large scales where it can artificially aid
explosion \citep{couch:15a,couch:14a}.

An accurate treatment of neutrino transport is crucial to simulating
CCSNe.  The neutrino mechanism hinges on how efficiently neutrinos can
move energy from where they decouple from the fluid near the
protoneutron star to just behind the shock
\citep[e.g.][]{janka:12a}. It appears that the success or failure of
3D CCSN explosion simulations is sensitive to the detailed properties
of the neutrino field.  In parameterized studies, increasing the
neutrino heating by just $\sim$$5\%$ can cause models to go
from failure to explosion \citep{ott:13a}. In models with more
realistic neutrino transport, small variations in the neutrino
opacities can mean the difference between success and failure
\citep{melson:15b}. Because of the strong energy dependence of weak
processes and the non-equilibrium nature of the neutrino field, CCSN
simulations require evolving the energy, and angle-dependent neutrino
distribution functions. In 3D time-dependent CCSN simulations, solving
the full Boltzmann equation is still computationally prohibitive (but
see \citealt{sumiyoshi:15} for static Boltzmann
solutions). To date, 3D radiation hydrodynamics simulations of CCSNe
have employed spectral, one-moment or two-moment radiation transport
schemes in the ray-by-ray approximation \citep[e.g.][]{lentz:15, melson:15b}.
Some argue that this approximation may overestimate spatial variations in the
neutrino field \citep[e.g.][]{sumiyoshi:15,skinner:16}.

General-relativistic (GR) gravity is another important ingredient in
CCSN simulations. Compared to simulations in Newtonian gravity, GR
simulations result in more compact protoneutron stars from which
neutrinos decouple at smaller radii and higher temperatures, resulting
in harder spectra. This effect appears to outweigh gravitational
redshift and leads to a higher neutrino heating efficiency.
\cite{mueller:12a}, in 2D, compared Newtonian, approximate GR, and
conformally-flat GR (exact in spherical symmetry) simulations for a
$15-$$M_\odot$ progenitor and found an explosion only in the GR
case. \cite{oconnor:15b} compared 2D Newtonian and approximate GR
simulations and also found GR effects to be essential for explosions.

In this paper, we present long term, fully 3D
radiation-hydrodynamics simulations of the postbounce phase of
CCSNe. Both hydrodynamics and neutrino radiation are evolved and
coupled on the same 3D grid. Our simulations are performed with the
\texttt{Zelmani} core collapse simulation package
\citep{ott:12a,ott:13a,reisswig:13a} that includes GR hydrodynamics
and GR spacetime evolution. For the first time, we use a new 3D
implementation of the GR spectral two-moment M1 approximation to
neutrino transport introduced by \cite{shibata:11}.

We carry out radiation-hydrodynamic simulations of the postbounce
evolution of a $27 \, M_\odot$ progenitor star in full 3D and, for
comparison, in octant symmetry, restricting the flow to an octant of
the 3D cube. All simulations are run to $\gtrsim
 380\,\mathrm{ms}$ after core bounce and each simulation is carried out
at two resolutions to test the dependence of the outcome on numerical
resolution.

We find that the shock in the full 3D
model begins to run away at around 220 ms after bounce in our highest
resolution run, suggesting that this model will achieve an explosion.
A model run at half this resolution experiences shock runaway at
around the same time, but shock expansion is much more
rapid.  Imposing octant symmetry on the high resolution run prevents
shock runaway.  In the lower resolution simulation, octant symmetry
does not prevent shock runaway but does marginally reduce the shock expansion
rate relative to the full simulation.

The remainder of this paper is structured as follows. In
Section~\ref{sec:methods}, we describe our simulation approach, setup,
and inputs. We discuss our simulation results in
Sections~\ref{sec:results} and~\ref{sec:discussion} and conclude in
Section~\ref{sec:conclusions}.

\begin{figure*}[t] 
\includegraphics[width=1.0\textwidth]{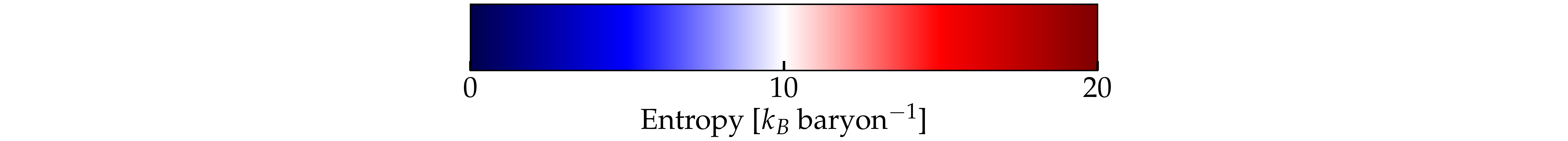} 
\includegraphics[width=1.0\textwidth]{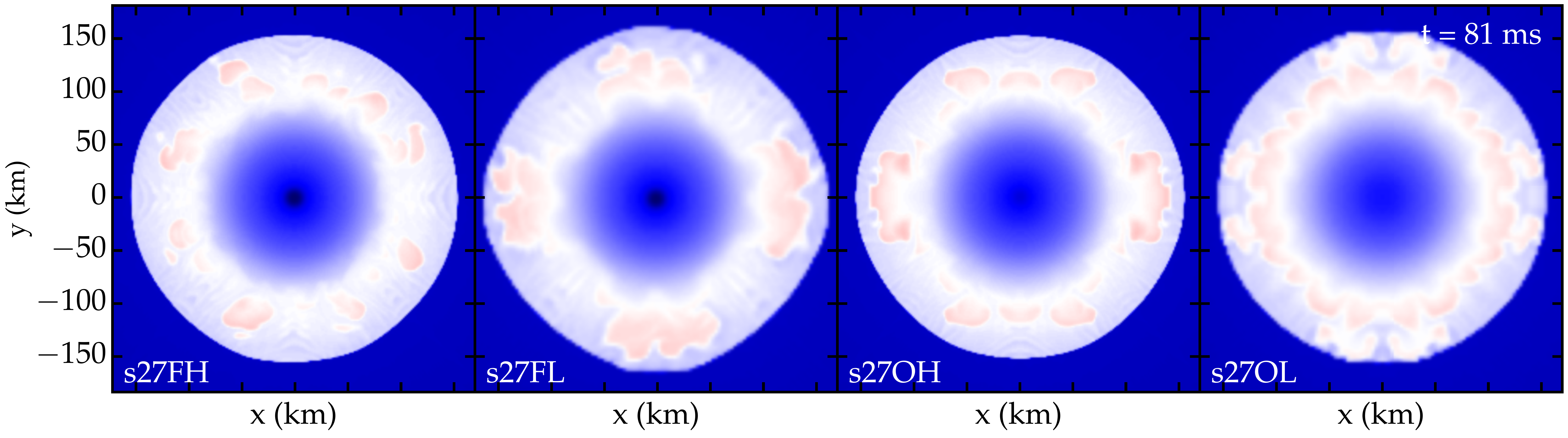} 
\includegraphics[width=1.0\textwidth]{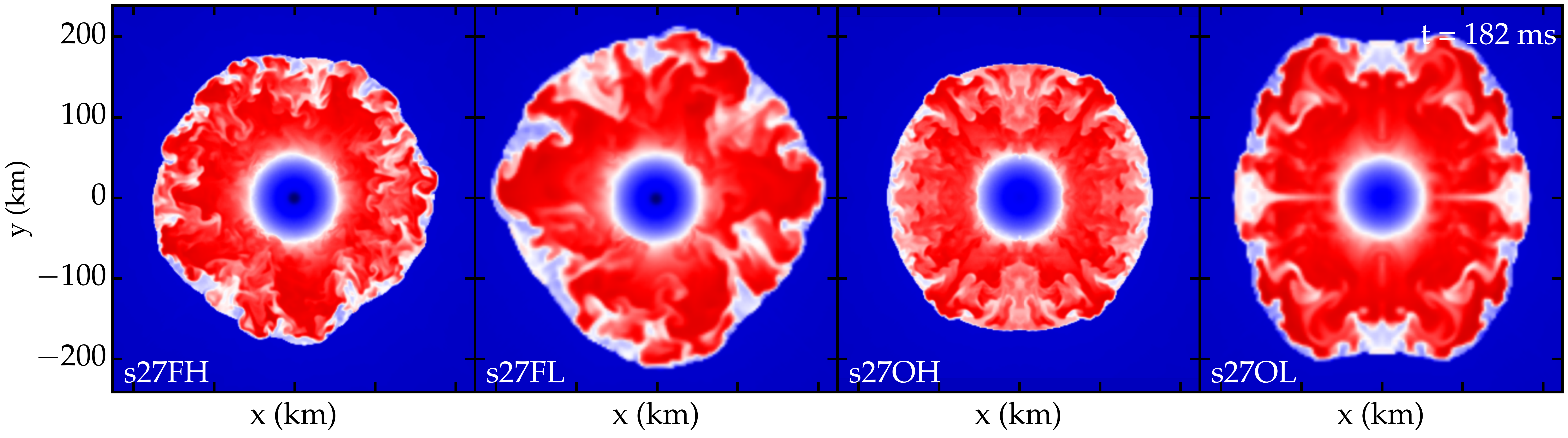} 
\includegraphics[width=1.0\textwidth]{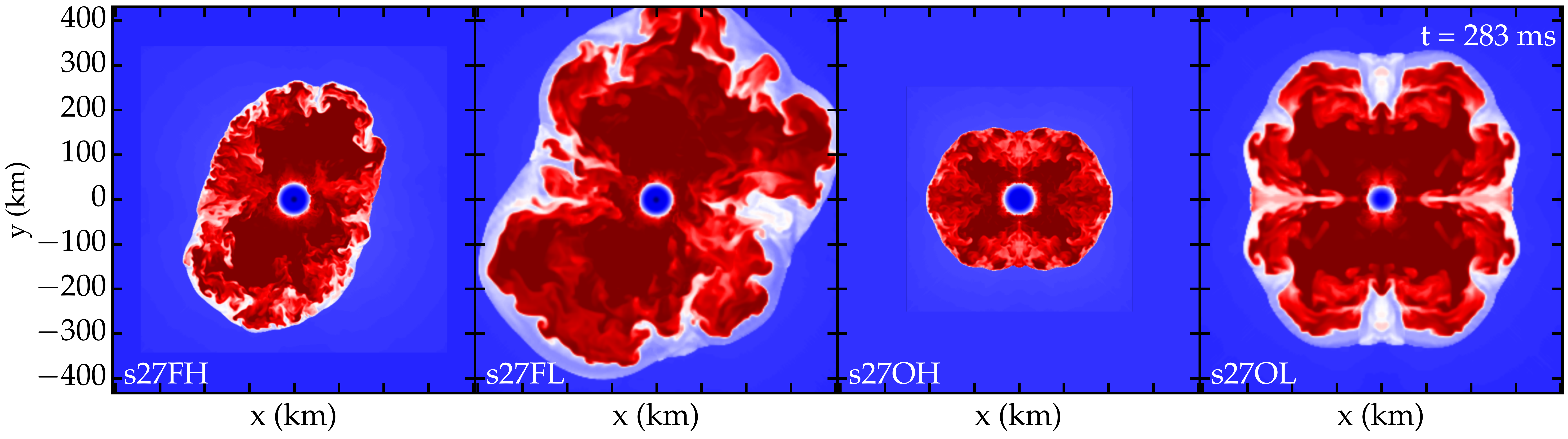} 
\includegraphics[width=1.0\textwidth]{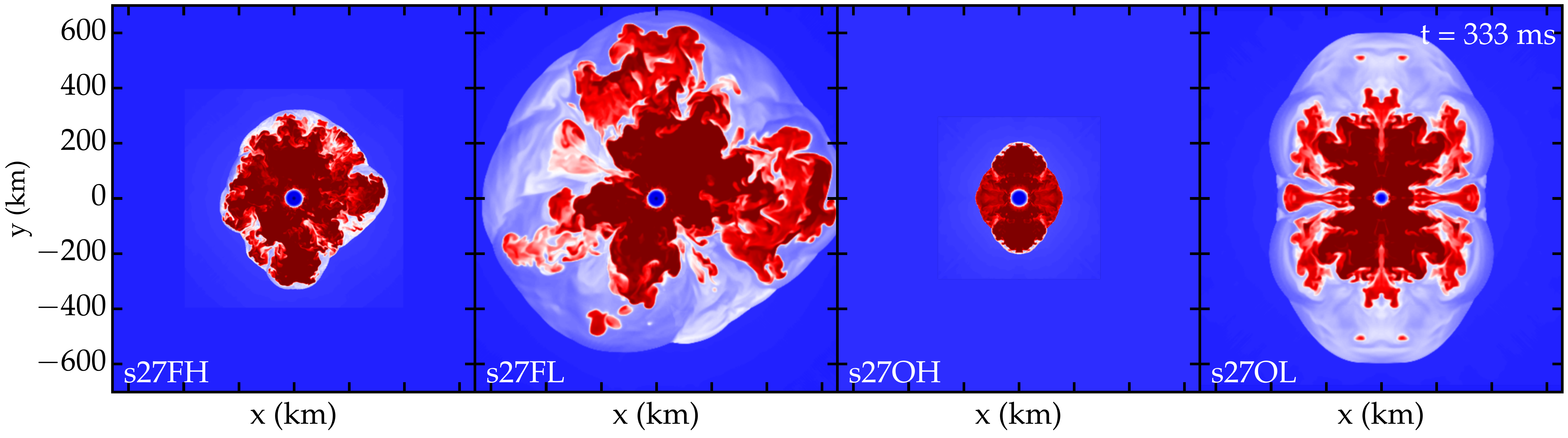} 
\caption{Cross-sections of the entropy distribution at selected times. The left
panels show the model \fullhi, the second from left panels show \fulllo, the
second from right panels show \octhi, and the right panels show \octlo. The
x-axis has the same scale as the y-axis. Notice that the spatial scale changes
in different rows but the entropy colormap stays constant.  The slight jump in
coloration in the accretion flow is artificial.  We only plot the refinement
level that includes the shock and outside of this region we choose a constant
background color to approximately match the coloration of the accretion flow.}
\label{fig:entropy}  
\end{figure*}

\section{Methods and Setup}
\label{sec:methods}

We employ the GR \texttt{Zelmani} CCSN
simulation package described in \cite{ott:12a,ott:13a} and
\cite{reisswig:13a}. \texttt{Zelmani} is based on the open-source
\texttt{Einstein Toolkit}~\citep{et:12,moesta:14a} and implements GR
hydrodynamics and spacetime evolution with adaptive mesh refinement
(AMR). See \cite{ott:13a,reisswig:13a,et:12} for implementation
details.

Within \texttt{Zelmani}, we have developed a multi-energy group GR 
M1 transport solver that evolves the radiation energy density $E_\nu$ 
and the radiation momentum density $F^i_\nu$ in the observer frame via 
the conservation equations 
\begin{eqnarray}
\partial_t \bar E + \partial_j 
\left(\alpha \bar F^j - \beta^j \bar E \right)
+ \partial_\nu \left(\nu \alpha n_\alpha \bar M^{\alpha \beta \lambda}
n_{\lambda;\beta}\right) \nonumber\\
= \alpha \left[\bar P^{ij}K_{ij} - \bar F^j \partial_j \ln \alpha
- \bar S^\alpha n_\alpha \right],
\end{eqnarray}
and 
\begin{eqnarray}
\partial_t \bar F_i + \partial_j
\left(\alpha \bar P^j_{\,\,i} - \beta^j \bar F_i \right)
- \partial_\nu \left(\nu \alpha \gamma_{i\alpha}
 \bar M^{\alpha \beta \lambda} n_{\lambda;\beta}\right)
\nonumber\\ = \bar F_k \partial_i \beta^k
 - \bar E \partial_i \alpha 
+ \alpha \frac{\bar P^{jk}}{2} \partial_i \gamma_{jk} +
\alpha \bar S^\alpha \gamma_{i\alpha} ,
\end{eqnarray}
where in the standard 3+1 GR notation, $\alpha$ is the lapse,
$\beta_i$ is the shift, $\gamma_{ij}$ is the three-metric, $K_{ij}$ is
the extrinsic curvature, $n^\alpha$ is the four-velocity of the
laboratory frame, $M^{\alpha \beta \lambda}$ is the third order
radiation moment (see \citealt{thorne:81}), $S^\alpha$ is the neutrino
source term, and $P^{ij}$ is the radiation momentum tensor. Over bars
denote densitized quantities, for example $\bar E =
\sqrt{\det(\gamma_{ij})}E$. To close this system of equations, we
assume $P^{ij} = P^{ij}(E_\nu, F^i_\nu)$ by interpolating between the optically 
thin and optically thick limits given in \citep{shibata:11}. \newtext{We employ the Minerbo
closure to interpolate between the optically thick and thin limits of
the radiation pressure tensor and third order radiation moment in the
fluid rest frame \citep{minerbo:78}. This is similar to the approach
discussed in \cite{shibata:11} and \cite{cardall:13rad}, and used in
\cite{just:15b}, \cite{oconnor:15b}, and \cite{kuroda:16}.  Our
numerical scheme is very similar to the gray scheme described in
\cite{foucart:15}.  The conservative moment equations shown above are
evolved using a finite-volume scheme, where the radiation quantities are
reconstructed at zone edges using the minmod limiter and we solve the resulting
Riemann problems approximately using the HLLE solver. The source terms are 
treated in a locally implicit fashion, while the red shifting terms are treated 
using an explicit finite volume upwind scheme.
 We evolve the velocity {\it independent} radiation
transport equations because we found numerical instabilities
to occur in the high optical depth limit when velocity dependence is
included.  Given that velocities behind the shock in the post-bounce
phase are small compared to the speed of light, this should be a
reasonable approximation.  We do not explicitly enforce lepton number
conservation.}

We draw the $27 \, M_\odot$ progenitor model s27 from \cite{whw:02},
which has been studied in a number of previous works
\citep[e.g.,][]{mueller:12b,hanke:13,ott:13a,couch:14a,abdikamalov:15}. In
all simulations, we employ the $K_0 = 220\,\mathrm{MeV}$ variant of
the equation of state of \cite{lseos:91} in the form described in
\cite{oconnor:10}. We follow collapse and the very early postbounce
phase in 1D using the open-source \texttt{GR1D} code
\citep{oconnor:15a,oconnor:13} without explicit velocity dependence
and the subset of \cite{bruenn:85} neutrino opacities laid out in
\cite{oconnor:13}, implemented via \texttt{NuLib} \citep{oconnor:15a}.
We use identical \texttt{NuLib} opacity tables in \texttt{GR1D} and
\texttt{Zelmani}, consider three neutrinos species ($\nu_e$,
$\bar{\nu}_e$, and $\nu_x =
       [\nu_\mu,\bar{\nu}_\mu,\nu_\tau,\bar{\nu}_\tau]$), and 12
       energy groups, spaced logarithmically with bin-center energies
       between $1\,\mathrm{MeV}$ and $248\,\mathrm{MeV}$.

We map to \texttt{Zelmani} at $30\,\mathrm{ms}$ after bounce and
continue the simulations in 3D with identical microphysics. For
mapping, we convert the \texttt{GR1D} metric to isotropic coordinates
and re-solve the Hamiltonian constraint in 1D (e.g.,
\citealt{baumgarte:10book}). \newtext{The neutrino fields are initialized to be in
equilibrium with the background fluid, which results in a short initial
transient in the luminosities and average neutrino energies.}

We carry out simulations in full 3D (``\fullhi'') without any symmetry
constraints and constrained ``octant'' 3D simulations (``\octhi''), in
which we simulate only in an octant of the 3D cube with reflective
boundaries on the $x-z$, $x-y$, and $y-z$ planes. Additionally, we
carry out lower-resolution simulations in full 3D and octant
3D, which we denote as ``\fulllo'' and ``\octlo,'' respectively.  Note
that our octant simulations differ from the rotational octant symmetry
employed, e.g., in \cite{ott:12a}, where periodic boundary conditions
are enforced on the $x-z$ and $y-z$ planes. This prevents us from
following any net rotation and likely changes the character of flows
near the boundaries.  This choice is made not for physical reasons,
but for computational savings.

All 3D simulations use Cartesian AMR with 8 levels of refinement.  We
do not employ the multiblock setup of \cite{reisswig:13a}. Each level
increases the resolution by a factor of two. The coarsest level
extends to $\sim$$6140\,\mathrm{km}$. We carry out simulations at two
resolutions.  In the \fullhi\ and \octhi\ simulations, the finest grid
covering the protoneutron star has a linear cell size of $\Delta x =
370\,\mathrm{m}$ and we use AMR to keep the entire postshock region
covered by the third-finest grid with $\Delta x = 1.48\,\mathrm{km}$
(corresponding to an angular resolution of $\sim$$0.85^\circ$ at a
radius of $100\,\mathrm{km}$). In the \fulllo, the finest linear cell
size is the same as in the high-resolution simulation, but we cover
the postshock region with the fourth-finest grid with $\Delta x =
2.96\,\mathrm{km}$ (corresponding to an angular resolution of
$\sim$$1.7^\circ$ at a radius of $100\,\mathrm{km}$). In \octlo, the
cell size on every refinement level is doubled relative to the
``high-resolution'' simulations and the shock is followed on the
third-finest grid with $\Delta x = 2.96\, \mathrm{km}$.

\section{Results}
\label{sec:results}

\begin{figure}[t] 
\includegraphics[width=1.0\columnwidth]{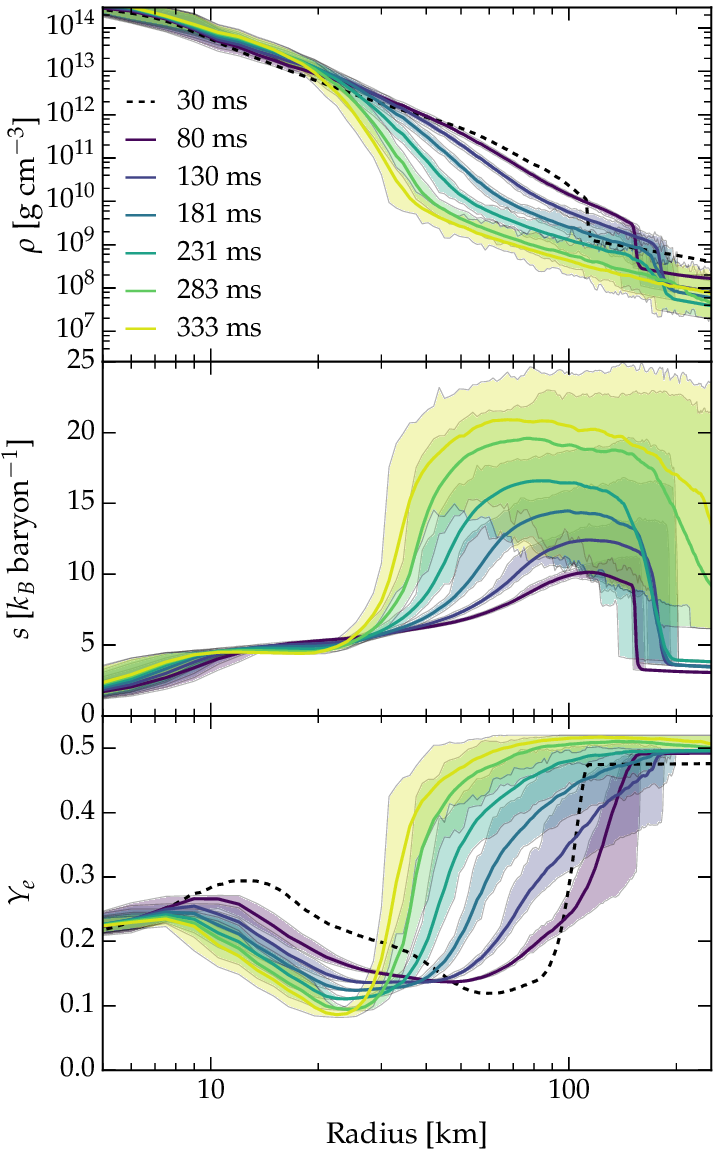} 
\caption{Evolution of the thermodynamic state of the gain region of
  \fullhi. The solid lines show the angle averaged density, entropy,
  and electron fraction of the ejecta, while the shaded regions show
  the minimum and maximum of these quantities on spherical shells. The dashed
  lines show the initial conditions of our 3D simulations.}
  \label{fig:structure}  
\end{figure}

We follow all four models for $\gtrsim$$380 \, \textrm{ms}$ after core
bounce or until the \SN{} shock has clearly run
away. Figure~\ref{fig:entropy} depicts entropy colormaps of equatorial
slices of all models at selected times. At late times, differences in
the numerical resolution and imposed symmetries can result in
qualitatively different evolution.  Both \fulllo\ and \octlo\ have
experienced shock runaway by $\sim$$280 \, \textrm{ms}$ and have
expanding high entropy regions and low entropy accretion streams
similar to what is seen in the simulations of \cite{lentz:15} and
\cite{melson:15b}, \newtext{although these other simulations employed different 
progenitors and had very different hydrodynamic evolutions}. In contrast, 
the \octhi\ shock has begun to
contract by $300 \, \textrm{ms}$ and does not contain large scale, coherent low entropy
downflows and high entropy outflows. The high-resolution, full 3D
simulation \fullhi\ has a continuously growing and deformed postshock
region, but does not run away as quickly as its low resolution
counterparts. Once again, the coherent low entropy accretion streams are less
prominent than the ones found in \fulllo. The evolutions of angle averaged thermodynamic
quantities in the postshock region of \fullhi\ are shown in
Figure~\ref{fig:structure}. The shaded regions in this figure indicate
angular variations that steadily grow with increasing postbounce time.

\begin{figure}[t] 
\includegraphics[width=1.0\columnwidth]{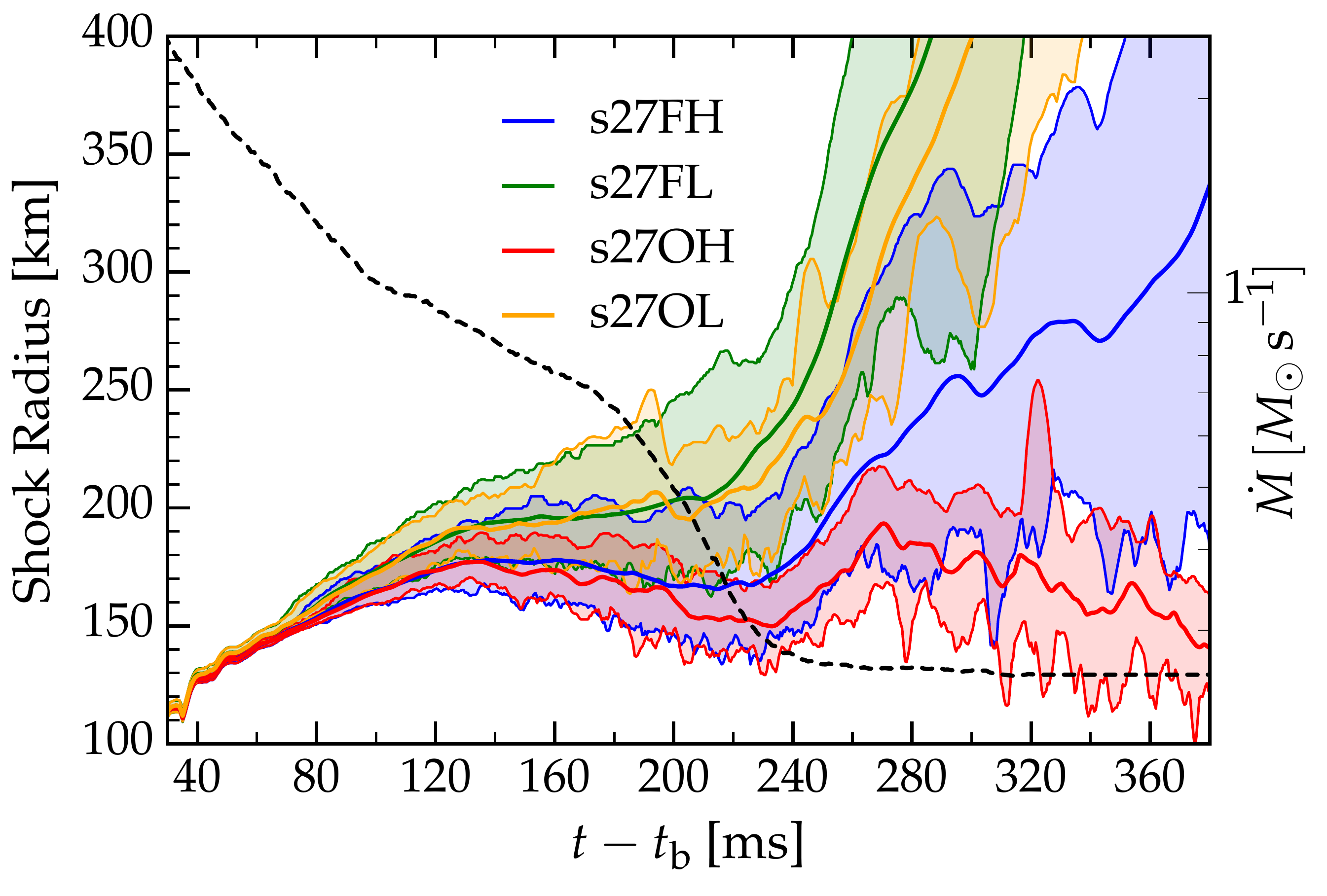} 
\caption{Evolution of the supernova shock. The thick lines show the
  angular average of the shock radius for all four models considered
  in this work.  For all four models, the
  shaded regions show the minimum and maximum radius of the shock at a
  given time. The dashed line shows the mass accretion rate in \octlo just
  outside the shock. The accretion rates for the other three models are
  similar.}
\label{fig:shock}  
\end{figure}

Our simulations are too computationally expensive to continue once the
supernova shock expands to large radii beyond
$\sim$$500\,\mathrm{km}$, since we use AMR to keep the entire
postshock region at constant resolution\footnote{When we stop it, the \fullhi\
simulation requires about $15\,\mathrm{TB}$ of main memory and is running on
$19,200$ NSF/NCSA Blue Waters CPU cores. The computational cost of this
model is approximately 60 million CPU hours.}. Therefore, we cannot follow the
evolution long enough to predict reliable explosion energies (or the diagnostic
energies considered in, e.g., \citealt{lentz:15}). Rapid shock expansion is our
best indicator of a possible explosion.  In Figure~\ref{fig:shock}, we present
the angle-averaged shock radius along with the angular variation of the shock's
position, indicated by shaded areas, bounded by minimum and maximum shock
radius. In all models, there are initially small oscillations as the model
relaxes after mapping from \texttt{GR1D}'s 1D spherical grid to
\texttt{Zelmani}'s 3D Cartesian AMR grid. Then the shock expands slowly and
secularly over the first $\sim$$120 \, \textrm{ms}$. The shock settles at
$\sim$$150-180\,\mathrm{km}$. In the high resolution models, the shock recedes
slightly and the differences between \fullhi\ and \octhi\ simulations are very
modest at this time. In the low resolution models, again independent of
full/octant, the shock maintains a nearly constant average radius for
$\sim$$80-100\,\mathrm{ms}$. The deviation of the minimum and maximum shock
radii from the average radius begin to increase over this period as the gain
region starts to convect in all simulations.  Around $230 \, \textrm{ms}$ after
bounce, the average shock radius begins to expand once again for all models. The
silicon-oxygen shell interface of the progenitor crosses the \SN{} shock at this time
and the accretion rate drops significantly (see Figure \ref{fig:shock}).
This is in agreement with the 3D simulation of \cite{hanke:13} (see their
Figure~2), however they did not find an explosion in 3D. In fact, the evolution
of the shock in our model \octhi\ is quite similar to the shock radius evolution
seen in \cite{hanke:13}. 

Clearly, the evolution of the shock depends significantly on both the resolution
of the simulation and on whether or not symmetries are imposed. As some of us
found in the parameterized 3D simulations of \cite{abdikamalov:15}, lower
resolution appears to be more favorable for shock runaway for simulations near
the threshold of explosion (cf.~\citealt{radice:16a}). At low resolution,
imposing octant symmetry does not have a large effect on the dynamics in the
gain region.  Both \fulllo\ and \octlo\ run away very quickly after the mass
accretion rate falls off, with \octlo\ lagging by only a few milliseconds.  The high-resolution full 3D simulation \fullhi\ runs
away more slowly than the low resolution simulations, but it nonetheless is
headed toward explosion, reaching a maximum shock radius of more than
$400\,\mathrm{km}$ and an average shock radius of $\sim$$315\,\mathrm{km}$ at
$370\,\mathrm{ms}$ after core bounce. The minimum shock radius of \fullhi\
barely expands after 260 ms, which is quite different from what is seen in the
low resolution models that experience rapid runaway in all directions. In the
octant high-resolution simulation \octhi, the shock begins to once again recede
soon after the passage of the silicon-oxygen shell interface. It seems very likely that
\octhi\ will result in a failed \SN{}, the two low resolution models are very
likely to explode, and \fullhi\ seems to be clearly on the path to explosion.

In Figure \ref{fig:shock_deformation}, we show the decomposition of the shock
front into real spherical harmonic modes following the convention in
\cite{burrows:12}, except that our spherical harmonic coefficients $a_{\ell,m}$ are larger by a factor of $\sqrt{2 \ell + 1}$. We present the root-mean-square amplitudes $A_\ell =
\sqrt{\sum_m a_{\ell m}^2}$ (where $a_{\ell,m}$ is a coefficient of the
spherical harmonic decomposition of $R_\mathrm{shock}(\theta, \phi)$). The top
two panels show the evolutions of the $\ell=1$ (for full 3D simulations) and
$\ell=2$ shock modes (for all models). For the high-resolution full-3D model
\fullhi, we show $\ell=1$ to $\ell=5$ in the bottom panel.  Considering an
expansion in a real spherical harmonic basis, our reflecting octant symmetry supresses odd
$\ell$ modes, all negative azimuthal modes, and odd azimuthal modes so that only
the $\{\ell=0,2,...;m=0,2,4,...,\ell\}$ modes can be excited.  This is in
contrast to rotating octant symmetry which allows for the modes
$\{\ell=0,2,...;m=0,\pm 4,\pm 8,\pm \ell\}$. This is very different from axial
symmetry (i.e.\ 2D simulations), where all of the $\ell$ modes can exist but all
$m$ modes except $m=0$ are supressed and small scale motions are effectively
constrained to two dimensions.

\begin{figure*}[t] 
\includegraphics[width=1.0\textwidth]{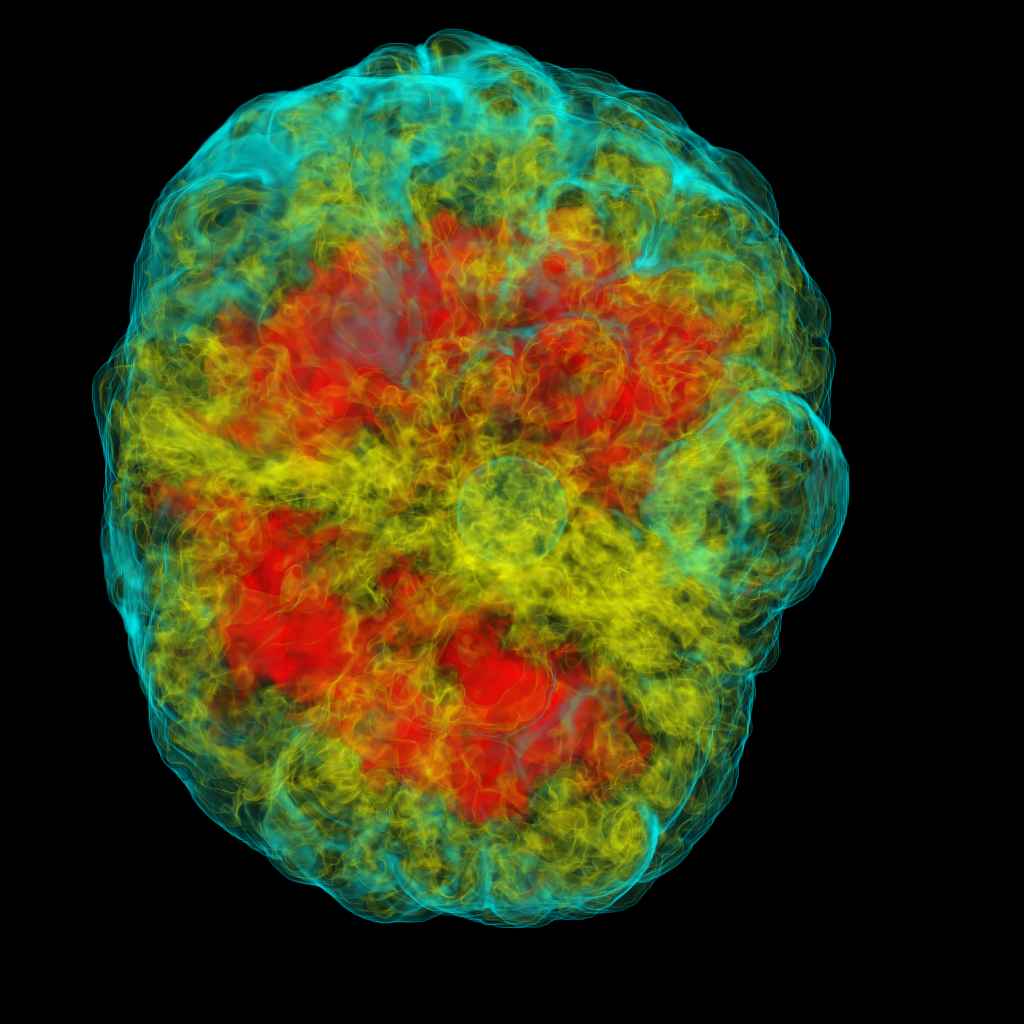} 
\caption{Volume rendering of the entropy distribution
    in the full 3D unconstrained high-resolution simulation
    \fullhi\ at $283\,\mathrm{ms}$ after core bounce. The cyan surface
    corresponds to the shock front and is at a specific entropy of
    $10 \, k_B\,\textrm{baryon}^{-1}$. The yellow regions are at
    specific entropies of $\sim 16 \, k_B\,\textrm{baryon}^{-1}$ and
    the red regions are at $\sim 20 \, k_B\,
    \textrm{baryon}^{-1}$. They correspond to strongly neutrino-heated
    bubbles of hot gas that expand, pushing the shock outward locally
    and globally.  This results in a complicated shock
    morphology that is asymmetric on large scale and on small scale.
    This figure was produced using {\tt yt} \citep{turk:11}.}
\label{fig:entropy_vol}  
\end{figure*}

All of the models experience increasing deviations from spherical symmetry with
increasing postbounce time. Although the asymmetry grows with time, none of the
models appear to be dominated by the standing accretion shock instability (SASI;
\citealt{blondin:03}). There is a period in \fullhi\ from $\sim 120 \,
\textrm{ms}$ to $\sim 240 \, \textrm{ms}$ where the $\ell=1$ mode oscillates
with constant frequency and grows, which may be indicative of SASI activity.
Nevertheless, these coherent oscillations are destroyed once the Si shell
interface is accreted through the shock.  Additionally, higher $\ell$ modes seem
to grow at the same rate. It is possible that the growth of low-order asymetries
without coherent oscillation is due to the SASI (which predicts longer period
oscillations with increased neutrino heating; \citealt{yamasaki:07, scheck:08}),
but it appears more likely that this asymmetry is driven by convective
instability in the postshock region (see Figure~\ref{fig:entropy_vol}).  The
SASI has been observed in some models that use the same s27 progenitor model and
hydrodynamics code, but include only parameterized neutrino physics
\citep{ott:13a, abdikamalov:15}. Strong SASI activity only occurred in these
models when the parameterized neutrino heating rate was low and shock runaway
did not occur.  When the parameterized neutrino heating rate was higher,
neutrino-driven convection dominated and much longer period ($\sim 20 \,
\textrm{ms}$) quasi-oscillatory behavior was observed, similar to what we find
here.

In the two unconstrained full 3D simulations, the $\ell=1$ mode begins to grow
rapidly once shock runaway occurs. Comparing with Figure~\ref{fig:entropy}, we
see that the late time asymmetry is driven by large solid angle regions of high
entropy outflow and cold accretion streams that penetrate to near the
protoneutron star.  Both $\ell=1$ and $\ell=2$ asymmetry increase during the
late shock expansion period of \fullhi, although it appears that the $\ell=1$
deformation is running away more rapidly. While it is not completely clear that
the shock is running away in \fullhi, this increasingly asymmetric expansion is
similar to what is seen in \fulllo, which clearly experiences shock runaway.
There is also strong $\ell=2$ deformation in \octlo\ after runaway. Although
\octhi\ does not experience shock runaway, it shows continued growth of the
$\ell=2$ and exhibits violent oscillations in the magnitude of the shock
deformation. This may indicate that an $\ell=2$ SASI is occurring in this model,
although the flow is not well ordered and it is hard to unambiguously determine
the contribution of convection relative to SASI.

\begin{figure}[t] 
\includegraphics[width=1.0\columnwidth]{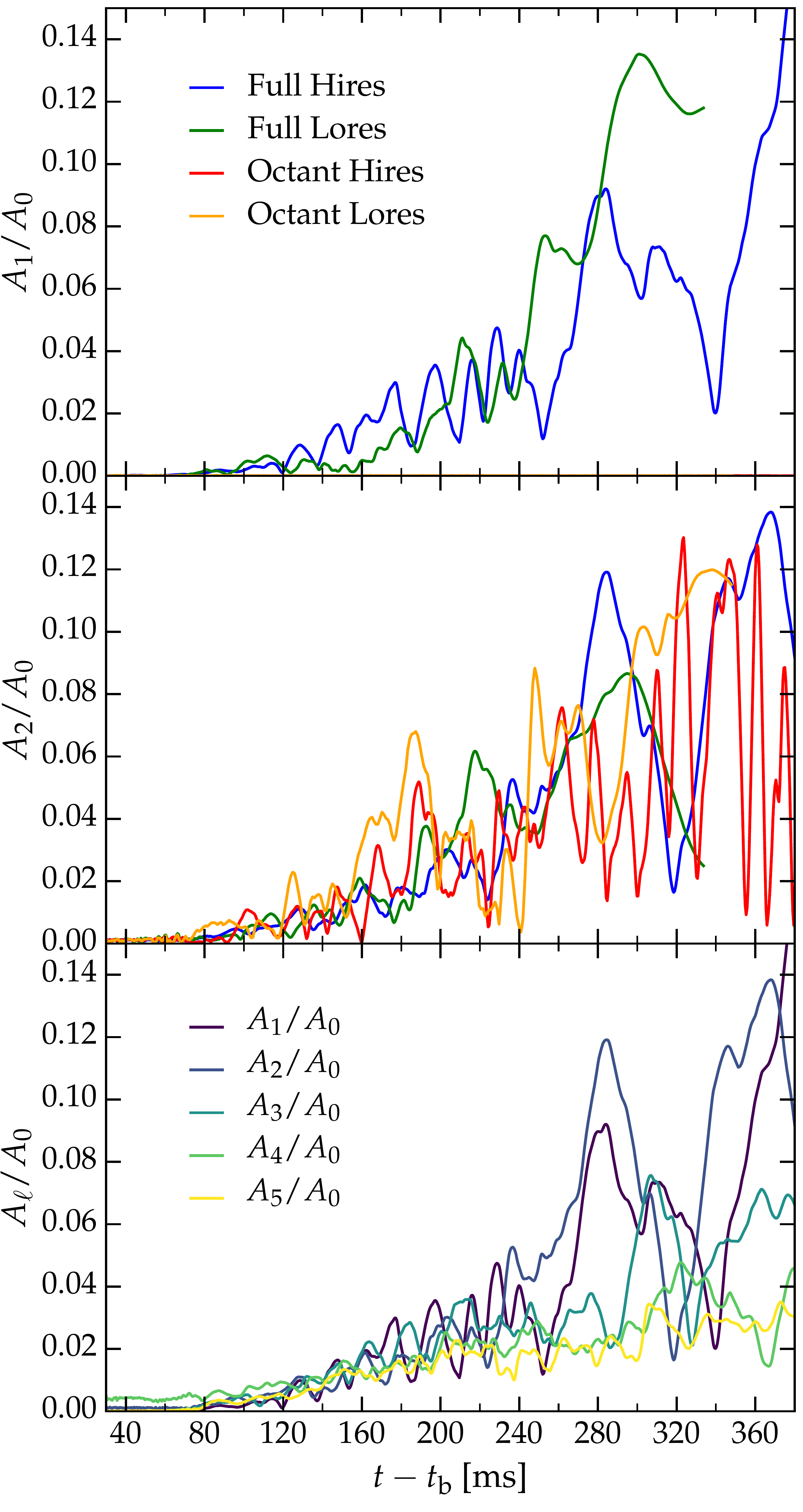} 
\caption{Evolution of the real spherical harmonic deformation of the supernova
shock front. {\it{Top Panel:}} The rms $m$ modes of the $\ell=1$ spherical
harmonic normalized to the $\ell=0, \, m=0$ mode. Octant symmetry forces all
$\ell=0$ modes to be zero. {\it {Middle Panel:}} Similar to the top panel,
except for the $\ell=2$ mode. {\it Bottom Panel:} Evolution of the first five
$\ell$-modes of \fullhi. }
\label{fig:shock_deformation}  
\end{figure}

\begin{figure}[t] 
\includegraphics[width=1.0\columnwidth]{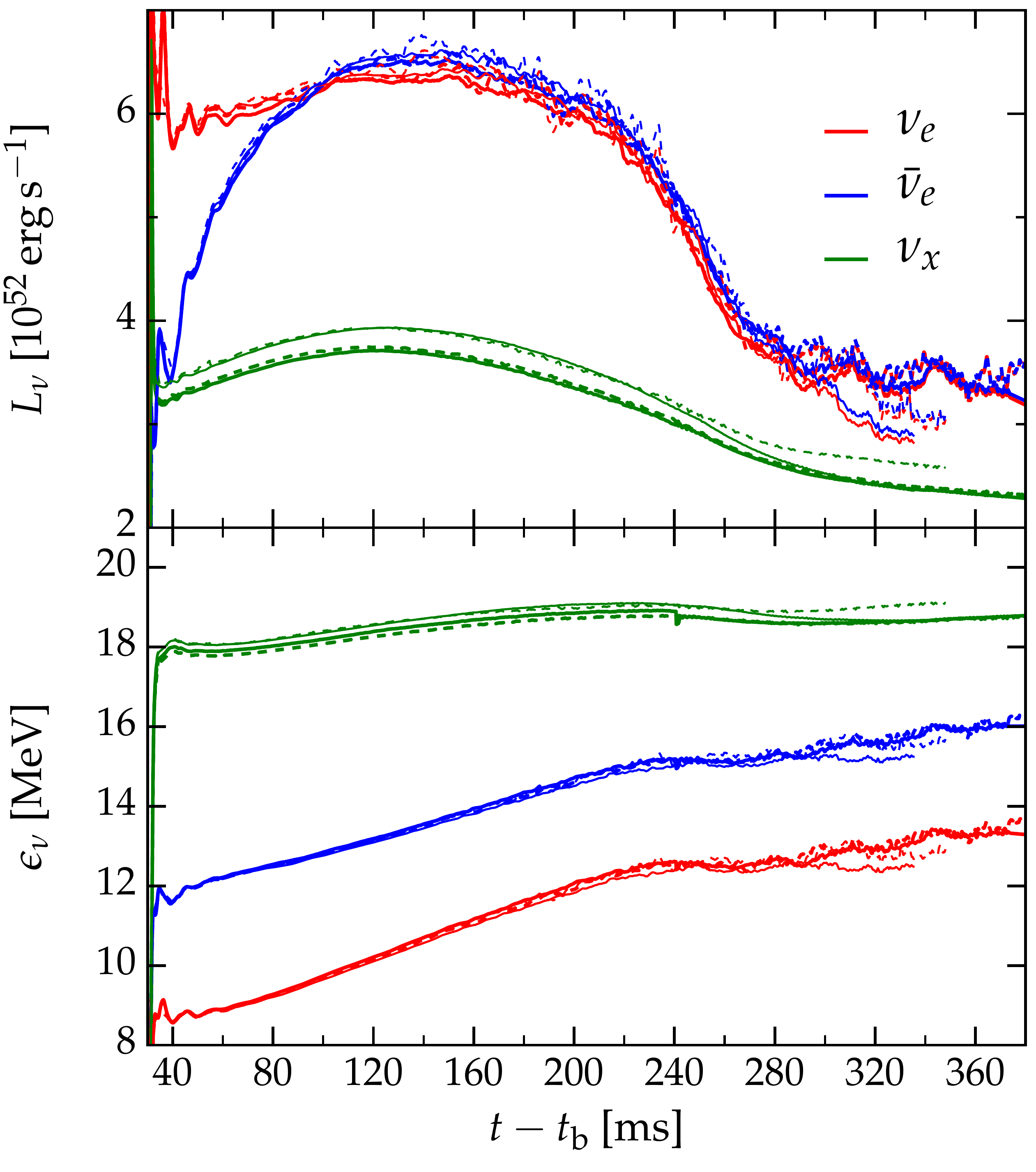}
\caption{Spherically averaged properties of the neutrino field at a
  radius of 450 km for the models \fullhi\ (thick solid lines),
  \fulllo\ (thin solid lines), \octhi\ (thick dashed lines), and
  \octlo\ (thin dashed lines).  The top panel shows the $\nu_e$ (red
  lines), $\bar \nu_e$ (blue lines), and $\nu_x$ (green lines)
  luminosities as functions of time.  The luminosities of \octhi\ are
  indistinguishable from those of the model \fullhi\ for the first
  $280 \, \textrm{ms}$.  The lower panel shows mean neutrino energies
  as a function of time.}
\label{fig:lums}  
\end{figure}

In Figure~\ref{fig:lums}, we show spherically averaged properties of the
neutrino field at a radius of $450\, \mathrm{km}$ for all four models.
Initially, there is a short period of oscillation in all quantities as the
initial spherically symmetric model relaxes on our 3D Cartesian grid.  These
oscillations cease by $\sim$$40\,\mathrm{ms}$ after bounce, and then the
spatially-averaged neutrino evolution is smooth.  Until $\sim$$280\,
\mathrm{ms}$ after bounce, there are only small differences between the neutrino
luminosities in all models.  Deviations after this time are due to large
variations in the extent and geometry of the postshock region and changes in the
accretion rate through the gain region (cf.~Figures~\ref{fig:entropy} and
\ref{fig:shock}).

All four models exhibit very similar average neutrino energies, the expected
hierarchy of neutrino energies, $\langle \epsilon_{\nu_e} \rangle < \langle
\epsilon_{\bar \nu_e} \rangle < \langle \epsilon_{\nu_{\mu/\tau}} \rangle$, and spectral
hardening as a function of time.  The large average energies of the
$\nu_{\mu/\tau}$, relative to the average energies predicted by other groups
\citep[e.g.][]{mueller:14}, are due to our neglect of inelastic neutrino
scattering. This is unlikely to have a large effect on heating in the gain
region, since $\mu$ and $\tau$ neutrinos do not effectively deposit their energy
there.  \newtext{ \citet{mueller:12a} have shown that inelastic scattering of heavy flavored neutrinos
near the electron antineutrino sphere can modestly increase the average energies
and luminosity of
electron flavored antineutrinos (by at most 10\%). Therefore, the neglect of inelastic
scattering in our models may change our quantitative results, but is unlikely to
make a qualitative difference in the outcome of our
simulations.  The slight offset in the $\nu_x$ luminosities between the low and
high resolution runs is due to different resolution near their neutrinosphere.} \cite{tamborra:14a} have also investigated 3D models of CCSNe using
the s27 progenitor.  Our $\nu_e$ and $\bar \nu_e$ luminosities and average
energies are within 10\% of those found by \cite{tamborra:14a}, but our
simulations show a different hierarchy of luminosities than theirs, with
$L_{\nu_e}<L_{\bar \nu_e}$.  Our $\nu_{\mu/\tau}$ luminosities are also about
25\% lower than those reported in \cite{tamborra:14a}. 

Additionally, \cite{tamborra:14} found that the lepton flux is
asymmetric about the center of mass with a strong dipole component,
i.e.\ their models exhibit a lepton emission self-sustained asymmetry
(LESA). In model \fullhi, we find that the dipole moment of the lepton
flux is less than $10\%$ of the monopole term at $280 \, \textrm{ms}$
after bounce. Previous to and after that time, it is even smaller.
Therefore, we do not see strong evidence for LESA in our highest
resolution model.  Conversely, \cite{tamborra:14} find a dipole moment
of the same order as the monopole moment at $280 \, \textrm{ms}$ after
bounce when using the same progenitor model.

There are a number of possible reasons for this discrepancy. First, it
has been suggested that the formation of LESA is related to
protoneutron star convection \citep{tamborra:14}. In \fullhi, we see
protoneutron star convection begin to develop only $\sim 230 \,
\textrm{ms}$ after bounce and it becomes fully developed only by $\sim
280 \, \textrm{ms}$.  The late onset of protoneutron star convection
is possibly due to the entropy and lepton number gradients in our
initial postbounce model, which did not include velocity dependence
and inelastic neutrino scattering during collapse.  The neglect of
these effects can significantly impact gradients of entropy and lepton
number inside the gain radius \citep{lentz:12a}.  Additionally, we
employ a set of neutrino opacities that differ in detail from the
opacities used by \cite{tamborra:14}, which can result in a different
evolution of entropy and lepton number gradients. It is also
possible that the full 3D neutrino transport we employ, as opposed to
the ``ray-by-ray'' approximation used by \cite{tamborra:14}, washes
out asymmetries in the neutrino field that drive the LESA
\citep{skinner:16,sumiyoshi:15}.  We emphasize that there are many
other differences between our neutrino transport scheme and the scheme
used by \cite{tamborra:14}, so the absence of LESA in our models
cannot be unequivocally attributed to the difference between full 3D
transport and the ``ray-by-ray'' approximation.

\section{Discussion} \label{sec:discussion}

In view of the small variations in neutrino properties between models, the
results of the previous section suggest that the effect of resolution and
symmetries on the postshock hydrodynamics, and consequently the shock radius
evolution, are of paramount importance.  The large variation of shock evolution
with resolution suggests that the post-shock hydrodynamics in our models are
unconverged \citep{radice:16a}.  Although our highest resolution simulation is
the highest resolution unconstrained neutrino radiation-hydrodynamics simulation
performed to date, it is still likely to be severely under resolved. In
\cite{abdikamalov:15}, the effective Reynolds number due to numerical viscosity
in simulations at the resolution employed here was estimated to be around 70.
This is many orders of magnitude lower than the physical Reynolds number in
these systems (although there is not a one-to-one correspondence between
physical and numerical viscosity; \citealt{radice:15a}).  Clearly, convectively
driven turbulence will behave differently at this low Reynolds number relative
to what would happen at the physical Reynolds number \citep{abdikamalov:15,
radice:16a}. \cite{abdikamalov:15} suggested that altering the resolution
changes the numerical viscosity and alters the spectrum of turbulence. It is
also possible that coarser Cartesian grids provide larger perturbations from
which turbulent convection can grow \citep{ott:13a}.  The size of the initial
perturbations is important since they must grow to macroscopic scales and become
buoyant before being advected out of the convectively unstable region
\citep{foglizzo:06,scheck:08}. 
 
The difference between unconstrained simulations and simulations enforcing
cylindrical symmetry has also been studied extensively, both with parameterized
or simplified neutrino physics \citep{nordhaus:10, hanke:12, couch:13b,
dolence:13, handy:14, couch:14a} as well as in models employing realistic
neutrino transport \citep{lentz:15}.  \newtext{The combined result of these
previous works has been somewhat inconclusive, but on the whole they seem to
(artificially) favor explosions in axisymmetry over full 3D.}  Axisymmetry
suppresses $m \neq 0$ large scale modes and makes a fluid behave as it would in
two dimensions at small scales. Both of these effects are likely to be
important, since large scale modes are important to the SASI and small scale
turbulence behaves very differently in two dimensions than in three
\citep{kraichnan:67}. \newtext{In contrast, our octant simulations suppress
the large scale $\ell=1$ modes but still permit true 3D fluid dynamics at small 
scales.} A comparison of the shock evolutions of \octhi\ and \fullhi\
suggests that the suppression of large scale $\ell$-modes makes it more challenging for
shock runaway to occur, all other things being equal at small scales.

\begin{figure}[t] 
\includegraphics[width=1.0\columnwidth]{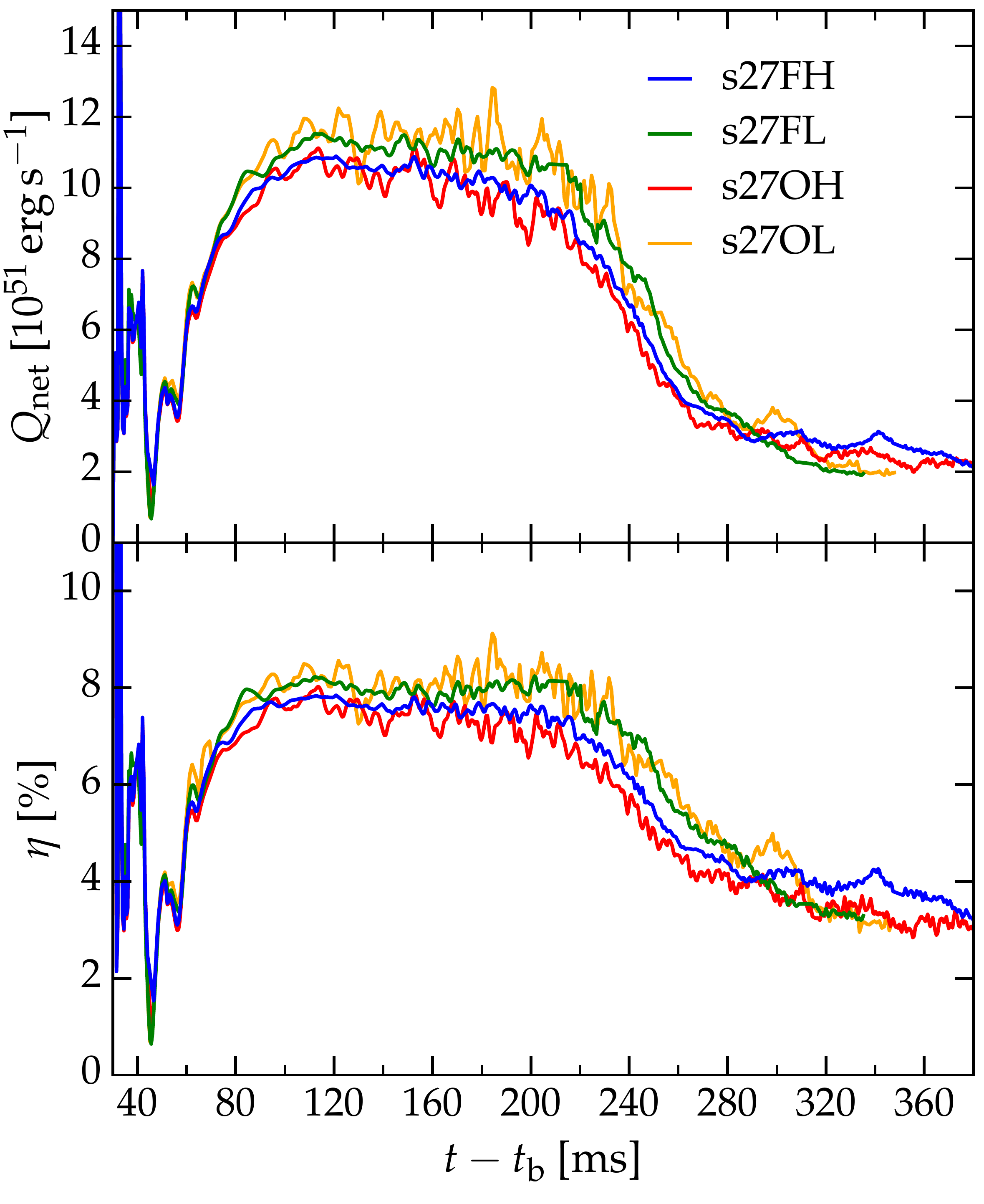} 
\caption{{\it Top panel:} The net neutrino heating rate in the gain region as a
function of time for \fullhi\ (blue), \octhi\ (red), \fulllo\ (green), and
\octlo\ (orange) averaged over a window of 2.5 ms. {\it Bottom Panel:} Heating
efficiency, $\eta = Q_{net}/(L_{\nu_e} + L_{\bar \nu_e})$, for these models.} 
\label{fig:heating}  
\end{figure}

The reasons for more rapid shock runaway at low resolution are less clear. We
find the properties of the neutrino field depend minimally on the resolution
(see Figure~\ref{fig:lums}). Therefore, the differences are unlikely to be due
to spatial resolution dependence of the neutrino transport. Nevertheless, it is
possible that differences in the structure of the gain region can result in
differences in neutrino heating.  The net heating rates and heating efficiencies
in the gain regions of the simulations are shown in Figure~\ref{fig:heating}.
We define the net heating rate $Q_\mathrm{net}$ as the integrated net neutrino
heating over regions that are experiencing net local heating.  The neutrino
heating efficiency $\eta$ is defined as the ratio of the net neutrino heating to
the sum of the electron neutrino and electron antineutrino luminosities just
below the gain radius. In the first $\sim$$75\,\mathrm{ms}$, there are minimal
differences between the four models. As the shock radii of the models begin to
diverge, the heating rates also diverge, with models with larger shock radii
experiencing larger heating rates and heating efficiencies. The models \fulllo\
and \octlo\ have similar averaged heating rates, although \octlo\ experiences
larger fluctuations once convection has developed. The average heating rate of
\fullhi\ is slightly larger than heating rate of \octhi\, the latter of which
also has a smaller average and maximum shock radius. \newtext{The neutrino
heating rate also shows non-radial variations due to non-radial variations in
the conditions of the fluid. If the net heating rate is decomposed into
spherical harmonics, the time evolution of the different modes is similar to the
evolution of the spherical harmonic modes of the shock decomposition. Therefore,
it seems likely that the differences between the simulations are due to
variations in the hydrodynamics, rather than the neutrino transport.} 

\begin{figure}[t] 
\includegraphics[width=\columnwidth]{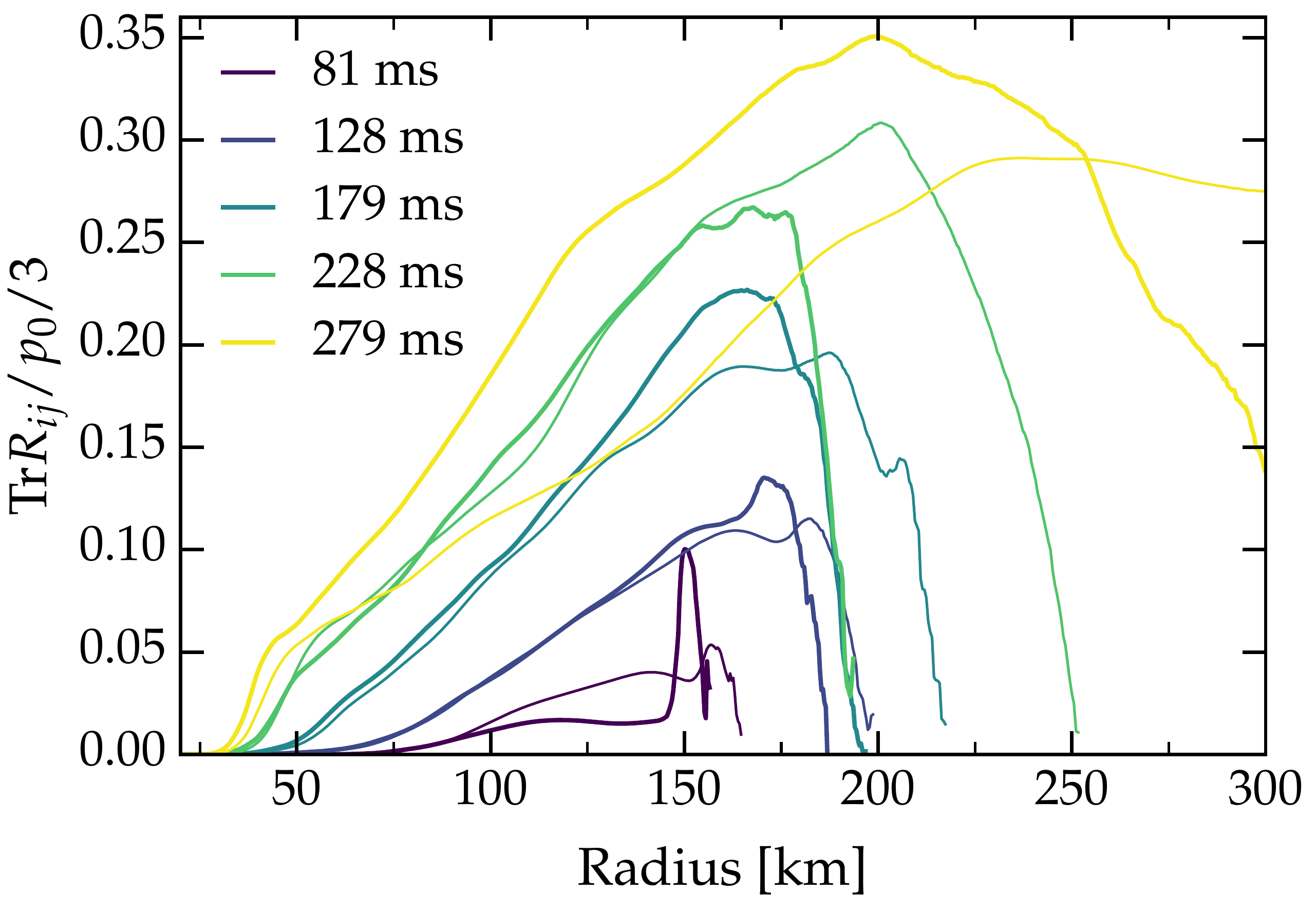}
\caption{The ratio of the Reynolds pressure (i.e. $\textrm{Tr}R_{ij}/3$) 
to the the average thermal pressure in the gain region.  The ratio as a function
of radius is shown at various times in \fullhi\ (thick lines) and \fulllo\ (thin
lines). Comparing with the $rr$ component of the Reynolds stress gives similar
results, although the maximum of $R_{rr}/p_0 \approx 0.5$.}
\label{fig:reynolds}  
\end{figure}

It is also possible that resolution may affect the properties of turbulence in
our simulations.  Therefore, we analyze our results in terms of the mean flow
equations \citep[e.g.][]{pope:00}.  The Reynolds stress can play a significant
role in the momentum equation in the gain region and behind the shock
\citep{murphy:11, murphy:13, handy:14, couch:15a, radice:16a}.  We denote the
Reynolds stress by $R_{ij} = \langle \rho v_i' v_j' \rangle $, where primes
denote fluctuations away from the mean.  $\textrm{Tr} R_{ij}/3$ acts like a
pressure in the averaged momentum equation and $\textrm{Tr} R_{ij}/2$ is the
kinetic energy contained in velocity fluctuations \citep{pope:00}. In Figure
\ref{fig:reynolds}, we show the ratio of the Reynolds pressure to the average
thermal pressure found in our simulations. Similar to \cite{couch:15a} and
\cite{radice:16a}, we find that the Reynolds pressure can be as large as a third
of the thermal pressure in a large portion of the gain region.  The maximum
contribution of the Reynolds stress is near the shock front.

From the perspective of the Reynolds decomposed Navier-Stokes equations, what
matters is the total energy contained in turbulent motions. The total energy is
directly related to the effective turbulent pressure, which can contribute
significantly to the total pressure in the postshock region and aid shock
expansion. \cite{abdikamalov:15} suggested that the resolution dependence of
CCSN simulations was due to differences in the spectrum of turbulence with
different effective numerical viscosity.  However, their Figure~13 shows little
resolution dependence in the turbulent kinetic energy at the large energy
containing scales that contribute most to the turbulent pressure.  Rather, they
find that resolution strongly affects the dissipation range, but the dissipation
range contains only a small fraction of the total turbulent kinetic energy at
any resolution.  This is consistent with the approximately equal Reynolds
stresses we see between in Figure~\ref{fig:reynolds}.

\begin{figure}[t] 
\includegraphics[width=\columnwidth]{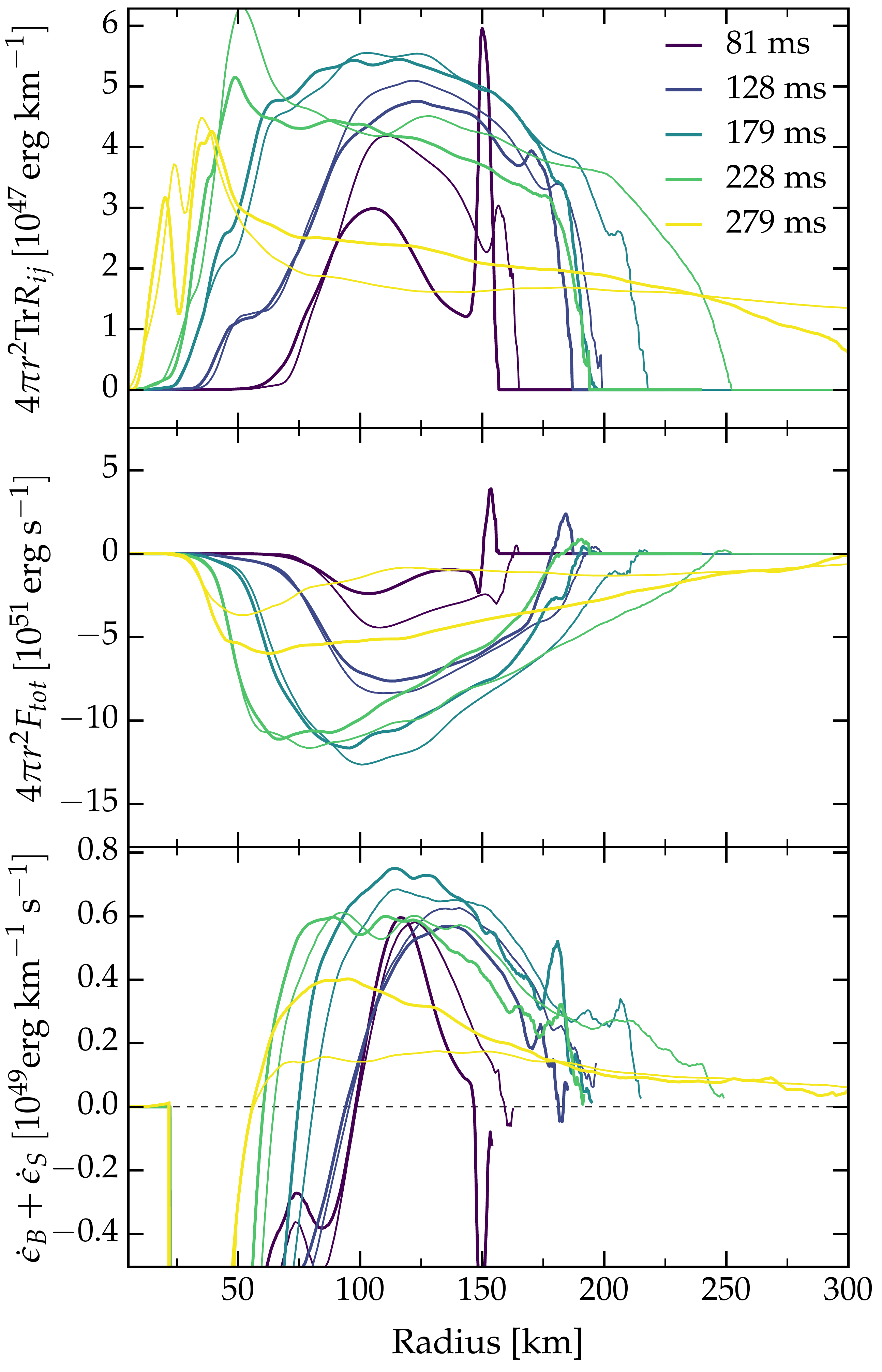}
\caption{The trace of the Reynolds stress and related quantities versus radius
at various times in \fullhi\ (thick lines) and \fulllo\ (thin lines). {\it Top
Panel:} The trace of the Reynolds stress integrated over angle. {\it Middle
Panel:} The total flux of the trace of the Reynolds stress. Throughout most of
the gain region, the flux is dominated by the average velocity advection term.
Near the shock, both pressure fluctuations and turbulent convection
significantly contribute. {\it Bottom Panel:} The Reynolds production by
bouyancy and shear.  Bouyancy is the dominant contributor throughout most of the
gain region, but the shear term contributes significantly near the shock.}
\label{fig:reynolds_transport}  
\end{figure}

Since the Reynolds stress contribution to the momentum equation can be a large
fraction of the contribution of the thermal pressure, it is instructive to
consider the evolution equation of the trace of the Reynolds stress. Including
compressibility and buoyancy effects, the trace of the Reynolds transport
equation is given by \citep{murphy:11}
\begin{equation}
\partial_t K + \partial_i ( v^i K + F_K^i + F_P^i) =
\dot \epsilon_S + \dot \epsilon_B 
+ \langle P' \partial_i v^i \rangle 
- \rho_0 \epsilon_\nu\,,
\end{equation}
where $K$ is the trace of the Reynolds stress, $F_K^i = \langle \rho v'_i
v'\cdot v'\rangle$ is the flux due to turbulent transport, $F_P = \langle P' v'
\rangle$ is the flux due to pressure fluctuations, and the terms of the right hand
side are the shear production term $\dot \epsilon_S = R_{ij} \partial_j v_i$,
the buoyancy production term $\dot \epsilon_B = \langle \rho' v_i' \rangle g^i$
($g^i$ is the gravitational acceleration), the work due to turbulent pressure
$\langle P' \partial_i v^i \rangle$, and viscous dissipation, $\rho_0
\epsilon_\nu$. The discussion here is mostly qualitative and we make no attempt
to include general relativistic effects. The latter are small anyway, since the
turbulent gain region is far away from the protoneutron star.

In Figure~\ref{fig:reynolds_transport}, we show the Reynolds stress and the
various terms that contribute to its evolution, integrated over angle, i.e. $4
\pi r^2 \langle \rho v^i v_i \rangle$, for \fullhi\ and \fulllo. We assume that
the average flow is spherically symmetric and calculate the average $\langle
\cdot \rangle$ over spherical shells. We neglect regions that lie outside of the
shock.  Although we do not plot it here, we find the well known result that
neutrino-driven turbulence is anisotropic on large scales with $R_{rr} \sim 2
R_{\theta \theta} \sim 2 R_{\phi \phi}$ \citep[e.g.,][]{murphy:13, couch:15a,
radice:16a}.  At all times before shock runaway, the Reynolds stress of \fulllo\ tracks the Reynolds
stress of \fullhi\ below $\sim$$100 \, \textrm{km}$.  The bottom panel of
Figure~\ref{fig:reynolds_transport} shows the net production of Reynolds stress
in the gain region. As was suggested by \cite{murphy:13}, buoyancy forces
provide the dominant contribution throughout most of the gain region. Although,
at some times, the shear production term can dominate the production just behind
the shock where the average fluid velocity is changing rapidly.  The production
is qualitatively similar in \fullhi\ and \fulllo, but it is systematically
higher at large radii in \fulllo. Throughout most of the gain region, the
Reynolds stress flux is inward and dominated by the $v^i K$ term.  Only near the
shock is there a small, outward flux of $K$. Once again, the high and low
resolution models are qualitatively similar.

We cannot easily extract the turbulent dissipation rate $\epsilon_\nu$ from our
simulations.  Nevertheless, we can infer some of its properties from differences
in the Reynolds stress with resolution. If the dissipation rate was very
sensitive to resolution, we would expect the Reynolds stress to saturate at
significantly different values when the resolution was changed. Rather, we find
that the Reynolds stress does not depend sensitively on resolution once
turbulence is fully developed. This result was also seen in the parameterized
simulations of \cite{radice:16a}, which extended to much higher resolutions than
we can consider here. 

During the period in which neutrino-driven convection is developing (i.e.\
before $\sim 100 \, \textrm{ms}$ after bounce), \fulllo\ has a Reynolds stress
that is a factor of $\sim 2$ larger than \fullhi\ from 100 to 150 km. This can
also be seen in the top row of Figure~\ref{fig:entropy}, where the lower
resolution models have convective plumes developing at slightly larger radii
than the higher resolution models.  This could potentially account for the
somewhat  more rapid shock expansion seen in \fulllo\ during the first
$\sim$$100 \, \textrm{ms}$ of evolution, although the extra contribution to the
pressure is at most only a few percent.  A plausible explanation for this
difference is that larger perturbations due to low resolution result in more
convective motion in model \fulllo\ at early times (cf.~\citealt{ott:13a}).
This may result in conditions more favorable for early time shock expansion,
which is consistent with the more rapid shock expansion seen in the low
resolution models at early times in Figure \ref{fig:shock}. This in turn results
in a somewhat larger neutrino heating rate and corresponding neutrino heating
efficiency, which makes conditions more favorable for eventual shock runaway.

At later times, when convection appears to be fully developed (see Figure
\ref{fig:entropy}), the variation of the Reynolds stress with resolution in the
gain region becomes smaller. At $130\,\mathrm{ms}$ and $180\,\mathrm{ms}$, $K$
is slightly larger in \fulllo\ above $100\,\mathrm{km}$.  \newtext{ Comparing with
Figure~\ref{fig:shock}, we see that the relative shock expansion rate is slightly larger in
\fulllo. At $230\, \mathrm{ms}$, around the time the shock radii of \fullhi\ and
\fulllo\ begin to diverge drastically, \fulllo\ has a much larger Reynolds stress
throughout most of the gain region.}  Nonetheless, the difference in the maximum
relative contribution to the pressure is only $\sim$5\% (see
Figure~\ref{fig:reynolds}).

Large grid perturbations are not expected in simulations that employ spherical
polar coordinates, but potentially large \emph{physical} perturbations are
expected from multi-dimensional stellar evolution simulations (e.g.,
\citealt{couch:15b}). In the abscence of physical or ad-hoc imposed
perturbations, the accretion flow is spherically symmetric and remains so on
spherical polar grids that typically induce much smaller numerical perturbations
into the flow than a Cartesian grid.  Although lower resolution was used in the
work of \cite{hanke:13}, their simulation using the s27 progenitor employed
spherical polar coordinates and their code is known to preserve spherical
symmetry. That model did not undergo shock runaway, while our models using the
same progenitor do.  Our results suggest that this qualitative difference may be
in part due to differences induced by the early strong development of
convection. Nevertheless, there are many other differences between our
simulation and theirs, so we caution against drawing definitive conclusions.
The results of \cite{radice:16a} are an important caveat. They carried out
parameterized neutrino-driven convection simulations over a large range of
resolutions.  Despite their use of perturbation-reducing spherical polar
coordinates, they found more rapid shock expansion at early times when the
resolution was reduced.

\section{Conclusions} \label{sec:conclusions} 

We have carried out fully 3D general-relativistic
multi-group neutrino radiation-hydrodynamics simulations of the postbounce phase
of core-collapse supernovae (CCSNe). We employed a $27$-$M_\odot$ progenitor
and followed its postbounce CCSN evolution for $380\,\mathrm{ms}$ at the highest
resolution to date. We observe the onset of explosion in low resolution and high
resolution full 3D simulations.

We find that both resolution and imposed large scale symmetries can have a
significant effect on the pre-explosion dynamics of CCSNe.  Shock runaway begins
in both of our fully 3D models at $\sim 230 \, \textrm{ms}$ after bounce, soon
after accretion of the silicon-oxygen shell interface.  While both models undergo shock
runaway, the lower resolution model runs away much more rapidly. The large
differences between the hydrodynamic evolution of these two models suggests that
at current resolutions, models of CCSNe are far from being converged, consistent
with the results found in parameterized studies \citep{abdikamalov:15,
radice:16a}. The imposition of octant symmetry in the high resolution model
prevents the shock from running away, while at low resolution octant symmetry
only has a modest effect on the gross features of the shock evolution. In
contrast to the hydrodynamic evolution, we find there are only small variations
in the properties of the neutrino field between simulations. \newtext{Our
results for the shock evolution of the s27 progenitor contrast with those of
\cite{hanke:13} and \cite{tamborra:14a}, who also performed 3D simulations and
found similar neutrino emission but did not observe shock runaway.}

In the models that experience shock runaway, the shock expansion is
asymmetric.  When no symmetries are imposed, the shock runaway in both
the high and low resolution simulations has a strong, growing $\ell=1$
deformation. In the octant simulation that experiences shock runaway,
there is a strong $\ell=2$ deformation, which is the lowest order
asymmetry available in octant symmetry.  Similar to what was seen in
the radiation-hydrodynamics 3D simulations of \cite{lentz:15}, we find
that this asymmetry is in part driven by coherent inflows and outflows
during shock runaway. The size of these structures is impacted by the
resolution. 

In previous work \citep{abdikamalov:15,radice:15a,radice:16a},
some of us argued that an inefficient turbulent cascade at low resolution traps
kinetic energy at large scales, artificially enabling shock expansion and
explosion. While it is true without doubt that kinetic energy at large scales is
what leads to shock expansion \citep[e.g.,][]{dolence:13}, the results of our
study suggest a more nuanced view on the resolution dependence of the neutrino
mechanism. We find some indication that lower resolution simulations have more
turbulent pressure support than higher resolution simulations. However, this is
true only at certain times and not universally throughout the postbounce
evolution. What may be equally or more important is how turbulent convection is
started: lower-resolution simulations seed turbulent convection with larger
numerical perturbations. This results in stronger turbulence early on that
pushes the shock out further and establishes a larger gain region, setting the
stage for a postbounce evolution that is more favorable for shock run away and
explosion.  It may thus be that Mazurek's law\footnote{Mazurek's law originated
in the context of stellar collapse at Stony Brook University in the 1980's when
Ted Mazurek was there. It is now used to generally refer to the strong feedback
in a complicated astrophysical situation which dampens the effect of a change in
any single parameter or condition (A.~Burrows and J.~Lattimer, \emph{private
communication}). } about the feedback-damping of perturbations applied to
complex nonlinear systems is violated after all: In critical cases, explosion or
no explosion may depend on the initial conditions from which turbulence grows.
This hypothesis clearly needs further scrutiny, but it falls in line with the
interpretation of (developing) turbulence as deterministic chaos
\citep{pope:00}.

The work presented in this paper and the conclusions that we draw have important
caveats and limitations. Much more future work is necessary to fully understand
neutrino-driven CCSNe. The most important limitations of our work are numerical
resolution in the hydrodynamic sector and the neglect of inelastic scattering
and velocity dependence in the neutrino sector. The latter two may significantly
effect the heating rate in the gain region and thereby the shock evolution.
Also, we started our simulations from a 1D postbounce configuration of a single
1D progenitor star. Future simulations should be fully 3D for the entire
evolution, consider a range of progenitors ideally coming from 3D presupernova
stellar evolution simulations, and should more conclusively explore the
resolution dependence of CCSN turbulence.

\section*{Acknowledgments}

The authors would like to thank E.~Abdikamalov, W.~D.~Arnett, A.~Burrows,
S.~Couch, F.~Foucart, K.~Kiuchi, J.~Lattimer, C.~Meakin, P.~M\"osta, D.~Radice,
Y.~Sekiguchi, and M.~Shibata for discussions. CDO wishes to thank the Yukawa
Institute for Theoretical Physics for hospitality during the completion of this
work. Support for LR during this work was provided by NASA through an Einstein
Postdoctoral Fellowship grant numbered PF3-140114 awarded by the Chandra X-ray
Center, which is operated by the Smithsonian Astrophysical Observatory for NASA
under contract NAS8-03060.  This research was partially supported by NSF grants
AST-1212170, CAREER PHY-1151197, PHY-1404569, AST-1333520, and OCI-0905046, the
Sherman Fairchild Foundation, and by the International Research Unit of Advanced
Future Studies, Kyoto University. Support for EO during this work was provided
by NASA through Hubble Fellowship grant \#51344.001-A awarded by the Space
Telescope Science Institute, which is operated by the Association of
Universities for Research in Astronomy, Inc., for NASA, under contract NAS
5-26555. This research was supported in part by Perimeter Institute for
Theoretical Physics. Research at Perimeter Institute is supported by the
Government of Canada through the Department of Innovation, Science and Economic
Development and by the Province of Ontario through the Ministry of Research and
Innovation. The simulations were carried out on the NSF XSEDE network
(allocation TG-PHY100033) and on NSF/NCSA BlueWaters (PRAC award ACI-1440083).
This paper has been assigned Yukawa Institute for Theoretical Physics report
number YITP-16-54.


\end{document}